\documentclass[letter,twocolumn]{jpsj2} %% for letters

\def\runauthor{Hisashi {\sc Kondo} and T\^{o}ru {\sc Moriya}}

\title
{
Superconductivity in Organic Compounds with Pseudo-Triangular Lattice
}

\author
{ 
Hisashi {\sc Kondo}$^{1,2}$ and T\^{o}ru {\sc Moriya}$^{3}$
}

\inst
{
$^{1}$Institute of Industrial Science, University of Tokyo, 
      4-6-1 Komaba, Meguro-ku, Tokyo 153-8505, Japan
\\
$^{2}$Computational Materials Science Center, National Institute for Materials Science (NIMS), 
      1-2-1 Sengen, Tsukuba, Ibaraki 305-0047, Japan
\\
$^{3}$Department of Physics, Faculty of Science and Technology, Tokyo University of Science, 
      2641 Yamazaki, Noda, Chiba 278-8510, Japan
}

\recdate
{December 12, 2003}

\abst
{
We study spin fluctuation (SF) mediated superconductivity (SC) in a half-filled square 
lattice Hubbard model with the transfer matrices $-t$ between nearest neighbor sites 
and $-t^\prime$ between a half of next nearest neighbor sites neighboring along only 
one of the $\langle 1,1 \rangle$ directions, considering application of this model to 
organic $\kappa$-(BEDT-TTF)$_2 X$ compounds. Varying the $t^\prime/t$ value from $0$ 
to $1$, one can interpolate between a square and an equilateral triangular lattice, 
the latter giving frustration to antiferromagnetically (AF) coupled spin systems. 
Within the fluctuation exchange (FLEX) approximation, we calculate 
$\chi(\mbox{\boldmath $q$},\omega)$, $T_{\rm c}$ and the SC order parameter for various 
model parameter values and find that both AF and SC are suppressed as one approaches 
the frustration geometry or $|(t^\prime/t)-1| \to  0$. The SC phase, however, extends 
beyond the AF phase boundary fairly close to $t^\prime/t = 1$ for realistic $U/t$ values. 
The order parameter is of $x^2-y^2$-type for $t^\prime/t < 1$ and of $xy$-type for 
$t^\prime/t > 1$.
}

\kword
{
organic superconductor, spin fluctuation, Hubbard model, triangular lattice, frustration
}

\begin{document}
\maketitle
%%%%%%%%%%%%%%%%%%%%%%%%%%%%%%%%%%%%%%%%%%%%%%%%%%%%%%%%%%%%%%%%%%%%%%%%%%%%%%%%%%%%%%%%

\begin{figure}[t]
  \begin{center}
    \includegraphics[scale=1.]{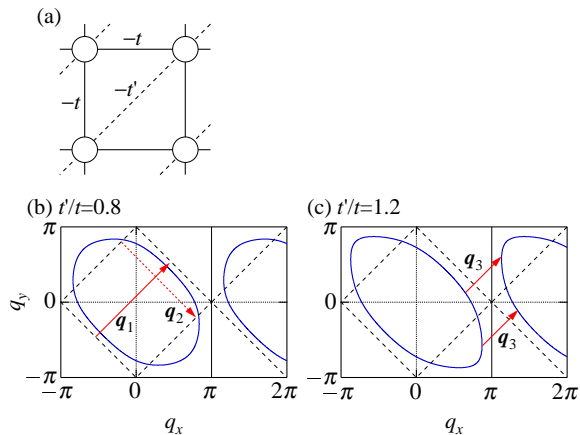}
  \end{center}
  \caption{
     (a) The model unit cell and the transfer integrals, and
     (b) and (c) unperturbed Fermi surfaces for $t^\prime/t = 0.8$ and $t^\prime/t =1.2$, 
     respectively. $\mbox{\boldmath $q$}_1 = (\pi,\pi)$, $\mbox{\boldmath $q$}_2 = (\pi,-\pi)$, 
     $\mbox{\boldmath $q$}_3 = (\pi/2,\pi/2)$.
  }
  \label{fig:1}
\end{figure}

According to recent investigations an organic compound $\kappa$-(BEDT-TTF)$_2$Cu$_2$(CN)$_3$ 
is a Mott insulator under ambient pressure showing spin liquid or resonating valence bond 
(RVB) behaviors without any magnetic order down to $32~{\rm mK}$.~\cite{rf:1} Under moderate 
pressure it undergoes a phase transition into a superconducting (SC) state with 
$T_{\rm c} = 3.9~{\rm K}$.~\cite{rf:2} The crystal structure of $\kappa$-(ET)$_2 X$, where 
ET stands for BEDT-TTF, is quasi-two dimensional, consisting of an approximate square lattice 
of dimers. It has been discussed that the square lattice Hubbard model with the transfer 
integrals $-t$ and $-t^\prime$ between antibonding dimer orbitals as shown in Fig. \ref{fig:1} 
well approximates the relevant electronic structures of this group of substances.~\cite{rf:3,rf:4} 
For $t^\prime/t = 1$ the model is topologically equivalent to the equilateral triangular lattice 
and the above-mentioned spin liquid like behaviors of $\kappa$-(ET)$_2$Cu$_2$(CN)$_3$ were 
interpreted in terms of the theoretical results for the Heisenberg model with a triangular 
lattice.~\cite{rf:5,rf:6} The estimated $t^\prime/t$ value for this substance is close to 
$1$ ($t^\prime/t \sim 1.1$), supporting this interpretation. 

As for the origin of superconductivity we first emphasize that the superconductivity occurs 
on the metallic side of the Mott transition and thus we are dealing with an intermediate 
coupling regime where a local magnetic moment is absent or not well defined. Furthermore, 
since the Mott transition is considered to be the first order phase transition, the SC phase 
is discontinuous with the insulator phase. Thus the SC phase may better be approached from 
the weak coupling or metallic side, with the consideration of electron-electron correlations. 
The situations are quite different from the doped Mott insulators in the strong coupling limit 
as is frequently represented by the $t$-$J$ model.~\cite{rf:7}

An equilateral triangular lattice is well known with its geometrical frustration for an 
antiferromagnetically (AF) coupled spin system occupying all lattice points. Although the 
ground state of the system is considered to be a $120$ degree spiral 
state~\cite{rf:8-1,rf:8-2,rf:8-3} the spin fluctuation is expected to have local singlet 
correlations, although without a spin excitation gap. For itinerant electron systems with a 
half-filled band in such a lattice with geometrical frustration, however, we rather expect to 
have multi- and broad peak structures of wave vector dependent susceptibility. The height of 
each peak is generally much lower than that of an antiferromagnetic peak in a lattice without 
frustration and thus the tendency toward magnetic ordering is strongly suppressed. 

Next we point out that the spin fluctuation (SF) mediated superconductivity occurs even when 
the system is fairly far from the magnetic instability insofar as the wave vector dependent 
part of the susceptibility is significantly enhanced. As a matter of fact we have demonstrated 
this situation in previous papers for the above-mentioned Hubbard model with 
$t^\prime/t \leq 1$ by using the fluctuation exchange (FLEX) 
approximation.~\cite{rf:9-1,rf:9-2,rf:9-3,rf:10} 

According to ref. \citen{rf:10} a calculated phase diagram in a $U/t$ against $t^\prime/t$ 
plane ($t^\prime/t < 1$) shows that the AF instability line tends to go upward or the critical 
value of ($U/t$)$_{\rm AF}$ tends to diverge as $t^\prime/t$ approaches $1$. Of course FLEX 
approximation is not reliable for very large values of $U/t$. Nevertheless it is interesting 
to see that this approximation clearly shows the absence of magnetic ordering in the lattice 
with frustration geometry. The SC instability line is located significantly lower than the AF 
instability line and the critical ($U/t$)$_{\rm SC}$ value increases rapidly as $t^\prime/t = 1$ 
is approached. However, the SC phase extends fairly close to $t^\prime/t = 1$ within realistic 
values for $U/t$ ($\sim 8$~\cite{rf:2,rf:3,rf:4}). $T_{\rm c}$ decreases as $t^\prime/t$ 
approaches $1$ and for $t^\prime/t = 1$ the SC phase is not found even for larger $U/t$ values.

The purpose of the present note is to report on the results of extended FLEX calculations for 
the same model with the values for the parameter $t^\prime/t$ around $1$ including the case 
of $t^\prime/t > 1$ and to discuss possible SF-induced superconductivity in this group of 
substances, including $\kappa$-(ET)$_2$Cu$_2$(CN)$_3$ under pressure. 

\begin{figure}[t]
  \begin{center}
    \includegraphics[scale=.95]{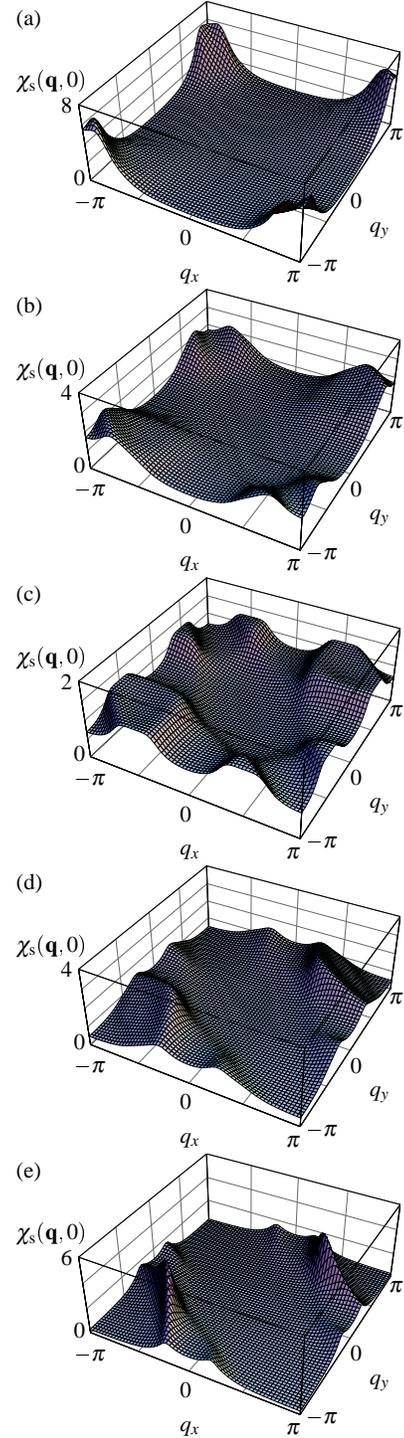}
  \end{center}
  \caption{
     Calculated wave vector dependence of the magnetic susceptibility for various parameter 
     values: (a) $t^\prime/t = 0.8$, (b) $t^\prime/t = 0.9$, (c) $t^\prime/t = 1.0$,
     (d) $t^\prime/t = 1.1$ and (e) $t^\prime/t = 1.2$ at $T/t = 0.02$ and $U/t = 10$. 
  }
  \label{fig:2}
\end{figure}

The model and approximation are the same as those discussed in 
refs. \citen{rf:9-1,rf:9-2,rf:9-3,rf:10}. We show in Fig. \ref{fig:2} the wave vector dependent 
susceptibility $\chi(\mbox{\boldmath $q$},0)$ in the \mbox{\boldmath $q$}-space for various 
values of $t^\prime/t$ and $U/t$. The peak at $(\pi,\pi)$ for the square lattice 
($t^\prime/t = 0$) is well known. With increasing $t^\prime/t$ this peak persists up to a 
fairly large value of it and then splits into two, their separation increasing as 
$t^\prime/t = 1$ is approached. For $t^\prime/t = 1$ we observe broad humps centered roughly 
around $(2\pi/3,2\pi/3)$ and $(-2\pi/3,-2\pi/3)$, etc. With still increasing $t^\prime/t > 1$ 
we see fairly sharp ridges passing through $(\pi/2,\pi/2)$ and $(-\pi/2,-\pi/2)$.  

This result combined with the Fermi surface geometry as shown in Fig. \ref{fig:1} suggests 
different symmetries of the SC order parameters for $t^\prime/t < 1$ and $t^\prime/t > 1$, 
as will be discussed below. As for the critical ($U/t$)$_{\rm AF}$ value for the magnetic 
instability, we find that the values are significantly larger for $t^\prime/t > 1$ than the 
corresponding (same distance from $t^\prime/t = 1$) values for $t^\prime/t < 1$.   

\begin{figure}[t]
  \begin{center}
    \includegraphics[scale=1.]{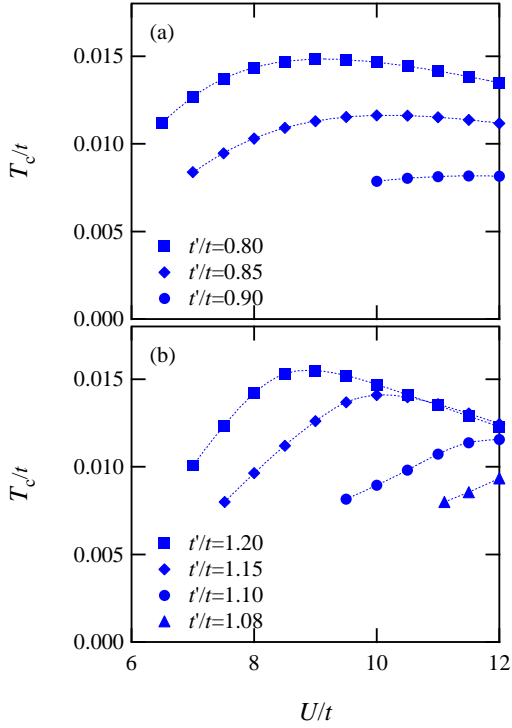}
  \end{center}
  \caption{
     Calculated values for $T_{\rm c}/t$ against $U/t$. (a) $t^\prime/t < 1$. (b) $t^\prime/t > 1$.
  }
  \label{fig:3}
\end{figure}

\begin{figure}[t]
  \begin{center}
    \includegraphics[scale=1.]{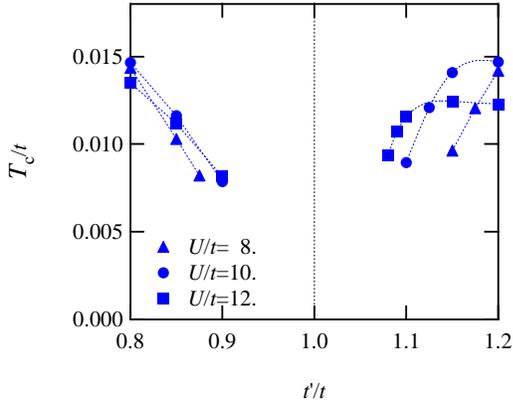}
  \end{center}
  \caption{
     Calculated values for $T_{\rm c}/t$ against $t^\prime/t$.
  }
  \label{fig:4}
\end{figure}

The calculated values for $T_{\rm c}/t$ against $U/t$ are shown in Fig. \ref{fig:3} for various 
$t^\prime/t$ values. In Fig. \ref{fig:4}, $T_{\rm c}/t$ is plotted against $t^\prime/t$ for 
several $U/t$ values.  We see that $T_{\rm c}$ decreases as $t^\prime/t$ approaches $1$. 
Although it is not easy to extend the present calculation to cover still lower temperatures or 
to convincingly extrapolate the results for $t^\prime/t > 1$ in Figs. \ref{fig:3} and \ref{fig:4} 
to lower values for $T_{\rm c}/t$, the present results clearly indicate that the SC phase is 
extended to cover a region fairly close to $t^\prime/t = 1$, including the location of 
$\kappa$-(ET)$_2$Cu$_2$(CN)$_3$  ($t^\prime/t \sim 1.1$). 

\begin{figure}[t]
  \begin{center}
    \includegraphics[scale=1.]{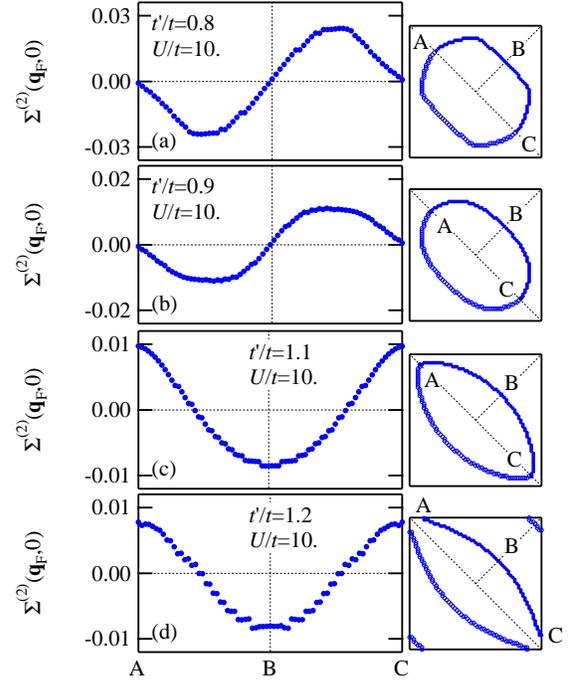}
  \end{center}
  \caption{
     Anisotropy of the SC order parameter showing $x^2-y^2$-type behavior for $t^\prime/t < 1$ 
     and $xy$-type behavior for $t^\prime/t > 1$ and corresponding Fermi surface geometry.
  }
  \label{fig:5}
\end{figure}

Some examples for the calculated \mbox{\boldmath $q$}-dependences of the order parameter are 
shown in Fig. \ref{fig:5} together with the corresponding Fermi surfaces. We see that the 
symmetry of the SC order parameter changes from $x^2-y^2$-type to $xy$-type as we go 
from $t^\prime/t < 1$ to $t^\prime/t > 1$.

This result is interesting in view of recent controversy as to the symmetry of the order 
parameter in $\kappa$-(ET)$_2 X$.  A recent study of the anisotropy of thermal conductivity 
in $\kappa$-(ET)$_2$Cu(CNS)$_2$ indicates that the symmetry of the order parameter is of 
$xy$-type.~\cite{rf:11} This is in contrast with the previous calculations indicating 
$x^2-y^2$-type for $t^\prime/t < 1$. Although $t^\prime/t < 1$ is predicted for this substance 
from theoretical estimations,~\cite{rf:4} approximations involved do not seem to be absolutely 
convincing. One might also expect some physical mechanisms to deform the Fermi surface geometry. 
In view of the above results of calculation systematic experimental studies for the SC order 
parameters and Fermi surfaces for a larger number of examples of $\kappa$-(ET)$_2 X$ are desirable.

As was seen from the present calculation as an approach from the weak coupling side, lattice 
structures with geometrical frustration are not favorable for SC, although the SC phase in the 
present example seems to extend with decreasing $T_{\rm c}$ fairly close to the equilateral 
triangular lattice. 

It is desirable to develop a theory for intermediate coupling regime interpolating between the 
strong and weak coupling limits. Such a theory is necessary in order to really understand the 
Mott transition. This seems to be a difficult problem particularly when we have a spin liquid 
state on the insulator side. With decreasing $U/t$ the range and strength of local singlet 
correlations in the spin liquid state are expected to change continuously until the M-I 
transition takes place. The possible importance or unimportance of singlet correlations on the 
metallic side of the Mott transition seems to be an even more difficult problem outside the 
scope of the present approach. We leave these problems for future investigations expecting the 
present approach to contain essential physics in this regime.

It should be noted that the present approach covers antiferromagnetic and frustrated regimes 
continuously with varying $t^\prime/t$ values and gives reasonable results for the former regime. 
For the latter regime we predict that with increasing geometrical frustration or $t^\prime/t$ 
approaching $1$, $T_{\rm c}$ is suppressed and for the triangular lattice, $t^\prime/t = 1$, 
we have no superconductivity. A part of the present predictions seems to be consistent with the 
experimental result for $\kappa$-(ET)$_2$Cu$_2$(CN)$_3$ as compared with the other 
$\kappa$-(ET)$_2 X$ superconductors. However, we have only one example of experimental result to 
compare in the spin liquid regime. Thus we had better wait for systematic experimental studies 
on various substances with $t^\prime/t$ close to $1$ before drawing a conclusion. 

We would like to thank Dr. H. Kino and Dr. S. Nakamura for useful discussions.

%%%%%%%%%%%%%%%%%%%%%%%%%%%%%%%%%%%%%%%%%%%%%%%%%%%%%%%%%%%%%%%%%%%%%%%%%%%%%%%%%%%%%%%%

%%%%%%%%%%%%%%%%%%%%%%%%%%%%%%%%%%%%%%%%%%%%%%%%%%%%%%%%%%%%%%%%%%%%%%%%%%%%%%%%%%%%%%%%
\end{document}